\documentclass[conference]{IEEEtran}

\usepackage{cite}
\usepackage{amsmath,amssymb,amsfonts}
\usepackage{algorithmic}
\usepackage{graphicx}
	\graphicspath{{Fig/}}
\usepackage{multirow}
\usepackage[caption=false, font=footnotesize]{subfig}
\usepackage{textcomp}
\usepackage{xcolor}
\def\BibTeX{{\rm B\kern-.05em{\sc i\kern-.025em b}\kern-.08em
    T\kern-.1667em\lower.7ex\hbox{E}\kern-.125emX}}

\makeatletter

\makeatletter
 \let\old@ps@headings\ps@headings
 \let\old@ps@IEEEtitlepagestyle\ps@IEEEtitlepagestyle
 \def\confheader#1{%
 \def\ps@headings{%
 \old@ps@headings%
 \def\@oddhead{\strut\hfill#1\hfill\strut}%
 \def\@evenhead{\strut\hfill#1\hfill\strut}%
 }%
 \def\ps@IEEEtitlepagestyle{%
 \old@ps@IEEEtitlepagestyle%
 \def\@oddhead{\strut\hfill#1\hfill\strut}%
 \def\@evenhead{\strut\hfill#1\hfill\strut}%
 }%
 \ps@headings%
 }
\makeatother

\begin{document}

\title{Unsupervised Learning Technique to Obtain the Coordinates of Wi-Fi Access Points
}

\author{\IEEEauthorblockN{Jeongsik Choi}
\IEEEauthorblockA{\textit{Intel Labs} \\
\textit{Intel Corporation}\\
Santa Clara, CA, USA \\
jeongsik.choi@intel.com}
\and
\IEEEauthorblockN{Yang-Seok Choi}
\IEEEauthorblockA{\textit{Intel Labs} \\
\textit{Intel Corporation}\\
Hillsboro, OR, USA \\
yang-seok.choi@intel.com}
\and
\IEEEauthorblockN{Shilpa Talwar}
\IEEEauthorblockA{\textit{Intel Labs} \\
\textit{Intel Corporation}\\
Santa Clara, CA, USA \\
shilpa.talwar@intel.com}
}

\maketitle

\begin{abstract}
Given that the accuracy of range-based positioning techniques generally increases with the number of available anchor nodes, it is important to secure more of these nodes.
To this end, this paper studies an unsupervised learning technique to obtain the coordinates of unknown nodes that coexist with anchor nodes. 
As users use the location services in an area of interests, the proposed method automatically discovers unknown nodes and estimates their coordinates.
In addition, this method learns an appropriate calibration curve to correct the distortion of raw distance measurements.
As such, the positioning accuracy can be greatly improved using more anchor nodes and well-calibrated distance measurements.
The performance of the proposed method was verified using commercial Wi-Fi devices in a practical indoor environment.
The experiment results show that the coordinates of unknown nodes and the calibration curve are simultaneously determined without any ground truth data.
\end{abstract}


\section{Introduction}
Wi-Fi is one of the most widely deployed wireless communication technologies.
In most indoor environments, a sufficient number of Wi-Fi access points (APs) are already installed to facilitate network connectivity. In addition to their original purpose, APs can be used as anchor nodes for estimating the location of mobile devices.
One simple approach for positioning is to measure distances from adjacent APs using received signal strength (RSS) and to apply trilateration techniques~\cite{Wang2003AnIW, 5425237, 5766644, 6488558}.
However, each environment has its own propagation characteristics, such as pathloss curve, which need to be investigated to achieve accurate positioning results.
Furthermore, RSS is also affected by many factors other than the distance from the transmitter~\cite{3gppSCM, 3gpp3DSCM}. Therefore, it is difficult to measure the exact distance from RSS.

To avoid using unreliable distance measurements in RSS, the Wi-Fi fingerprinting method has been widely studied as a range-free positioning technique~\cite{832252, 1047316, horus05, 6550414, 7438932}.
Instead of measuring distance based on RSS, this method prepares a database called a radio map, which tabulates RSS measurements from all the neighboring APs at the selected coordinates.
During the location estimation phase, the coordinates of the device can be obtained by identifying an entity in the pre-built radio map that has the closest RSS measurements, from the current measurements.
The fingerprinting method also requires time and effort to prepare a radio map for each environment.

The aforementioned Wi-Fi-based positioning techniques are based on a prior standard that was not originally designed for positioning purposes.
To enhance the positioning capability, the IEEE 802.11-2016 standard (also known as 802.11REVmc) introduced a new ranging protocol called fine timing measurement (FTM). This protocol measures the distance between two nodes based on the round trip time (RTT).
Although the performance of the FTM protocol still depends on many factors such as the presence of a line-of-sight (LOS) path, obstacles in the LOS path, the transmission bandwidth, hardware calibration, and so on, it potentially provides more accurate ranging results compared to RSS~\cite{intel17, Ibrahim:2018:VAE:3241539.3241555, choi19}.

Since FTM uses the flight time of radio waves for which the propagation speed is almost the same for all environments, the effort required to investigate the characteristics of each environment might be reduced.
The most time-consuming task is to obtain the coordinates of APs for use as anchor nodes.
For instance, if APs are installed in ceilings or in a restricted area, it is not easy to obtain their coordinates.
In addition, in the case that non-fixed APs are used as anchor nodes, their coordinates should be regularly updated.
More importantly, there may be a significant number of public APs deployed in an unplanned manner.
Once their coordinates are available, the positioning accuracy is improved as the number of anchor nodes increases.

\begin{figure*}
\subfloat[]{\includegraphics[width=0.32\textwidth]{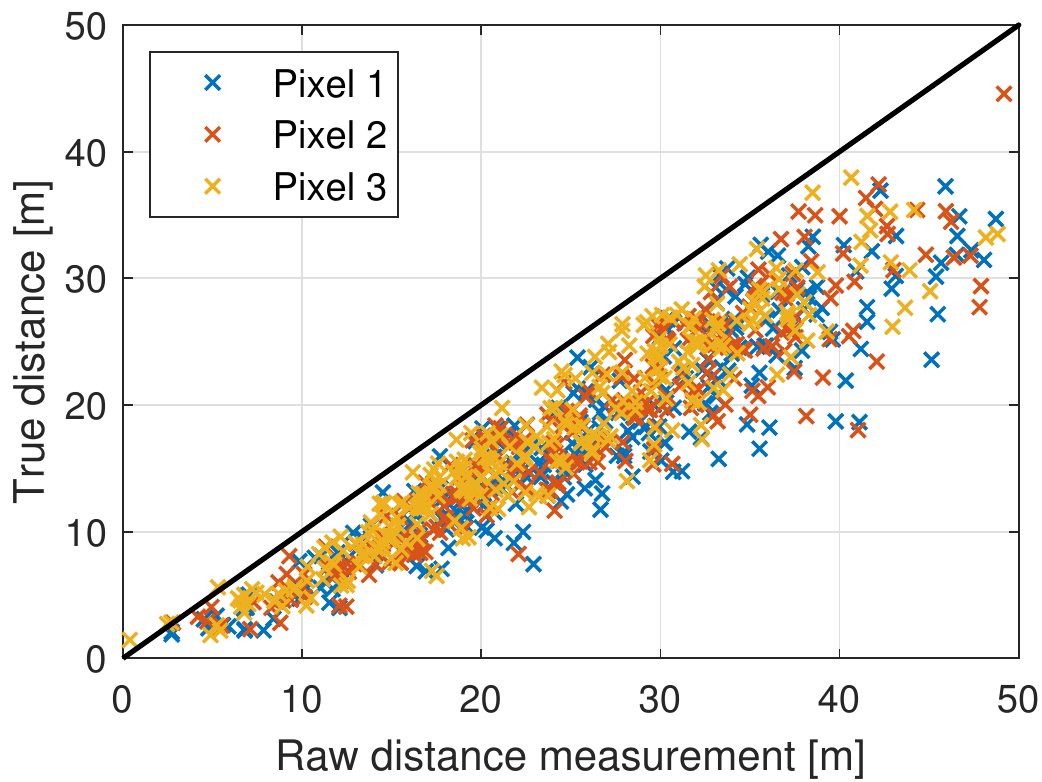}}\hfil
\subfloat[]{\includegraphics[width=0.32\textwidth]{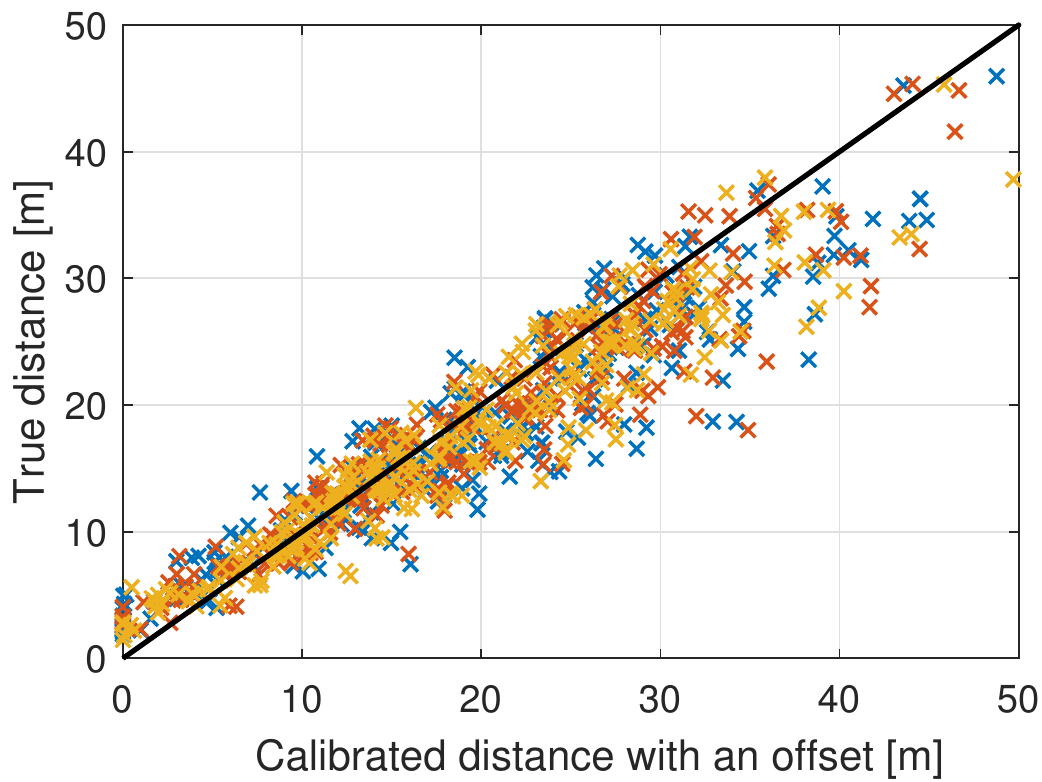}}\hfil
\subfloat[]{\includegraphics[width=0.32\textwidth]{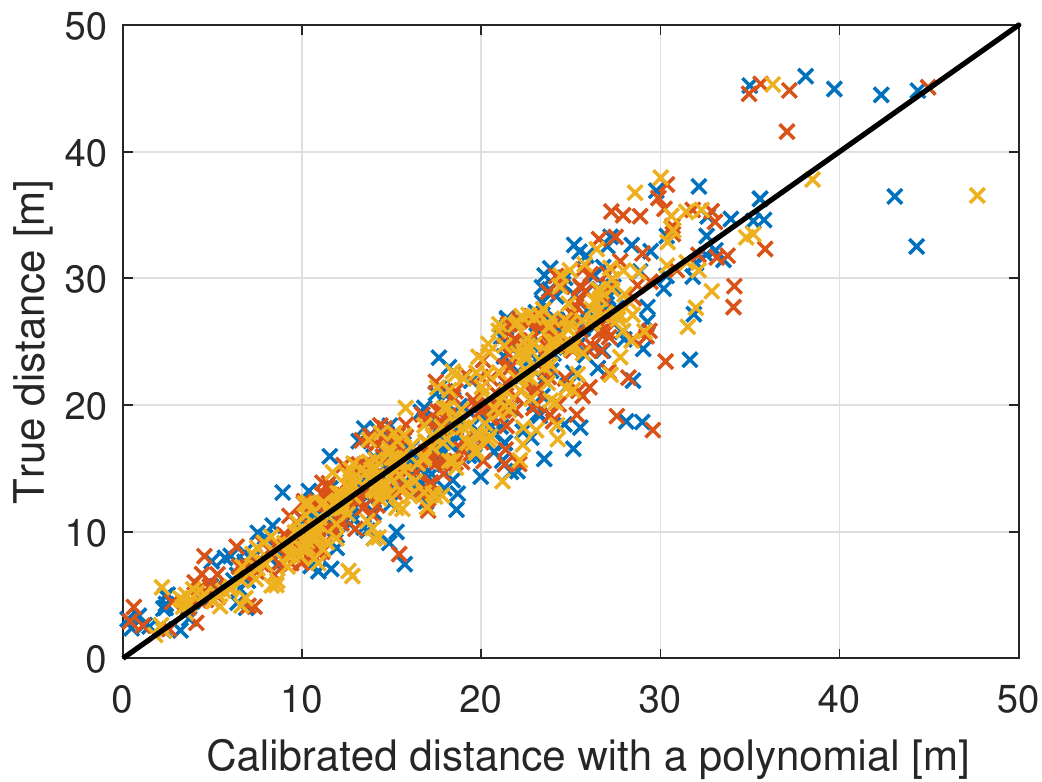}}
\caption{The relationship between the true distance and the followings: (a) raw distance measurement from the FTM protocol, (b) calibrated distance using an offset, and (c) calibrated distance using a second-order polynomial. The data were obtained using Google Pixel series (Pixel 1, 2, and 3) running on Android~9.}
\label{fig_ftm_scatter}
\end{figure*}

One approach for estimating the coordinates of unknown APs is to apply the trilateration technique using a mobile device as an anchor node~\cite{8053692}.
By measuring distances to unknown APs at multiple known positions, the coordinates of each unknown APs can be obtained.
This is a supervised method given that the coordinates of the device should be recorded whenever the measurement occurs.
Another approach is to apply multidimensional scaling (MDS), which has been widely studied in wireless sensor networks~\cite{Costa:2006:DWS:1138127.1138129, 8439937}.
This method estimates the coordinates of unknown nodes from the coordinates of a few anchor nodes using the measured distances between nodes.
To apply this method to Wi-Fi, it is necessary to change the configuration of some APs, because the current FTM protocol measures the distance between two nodes running in different modes.
Moreover, this method does not work well if some APs are isolated from others.

In this paper, we study an unsupervised learning approach to obtain the coordinates of unknown APs that coexist with anchor APs. 
As long as users simply use the location service at the area of interest, the proposed method automatically discovers unknown APs and estimates their coordinates.
Moreover, given that users can go anywhere in the area and connect multiple APs in the middle, the proposed method can be applied even if some unknown APs are isolated from others.
To this end, the cost functions introduced in~\cite{choi19} are used to evaluate the geometric validity of the relationship between the coordinates of anchor nodes and distance measurements.

In addition to estimate the coordinates of unknown APs, we also focus on the calibration of the FTM protocol.
According to the experiment results obtained in practical environments~\cite{Ibrahim:2018:VAE:3241539.3241555, choi19}, raw distance measurements using the FTM protocol yield biased or distorted results of the true distances.
To resolve this, a calibration procedure is introduced in~\cite{ftm_cal}.
By manually acquiring distance measurements at various distances from the AP, a calibration curve can be obtained.
However, the proposed method learns an appropriate calibration curve without any labeled data, even while estimating the coordinates of unknown APs.

The remainder of this paper is organized as follows. In the next section, we introduce the main assumptions and FTM measurement model.
In Section III, we define cost functions for estimating unknown parameters in the system, including the coordinates of unknown APs. In Section IV, the performance of the proposed method is evaluated in a practical environment using off-the-shelf APs and devices. The conclusions of the paper are summarized in Section V.

\section{System Model}

We consider a two-dimensional area with anchor APs and unknown APs installed.
We denote $\mathcal N$ as the set of all APs including both anchor and unknown APs, and $N$ as the number of elements in this set.
To collect training data, mobile devices move arbitrarily around the area of interest while periodically measuring distances from all the adjacent APs using the FTM protocol.
The biggest benefit of unsupervised learning is that the training data can be easily collected even when users are using the location services.

According to the experiment results in~\cite{Ibrahim:2018:VAE:3241539.3241555, choi19} and this work, the uncalibrated FTM protocol provides distorted distance measurements of the true distances.
This is due to the timing offset of the FTM packets, cable length, hardware offset, or even site-specific factors\footnote{The performance of the FTM protocol depends on how accurately detect the arrival time of the direct path. Therefore, it may be necessary to apply different calibration methods depending on the environment (e.g., free space and rich scattering environments).}.
Fig.~\ref{fig_ftm_scatter}(a) illustrates the relationship between true distances and raw distance measurements based on the FTM protocol. The details of the experiments are presented in Section~IV.
It is evident that there is a mismatch between the raw measurements and true distances.

Based on this observation, the distance measurements between the $n$-th AP and a device at time step~$i$ can be simply expressed as a function of true distance as follows:
\begin{equation} \label{2_1}
d^{FTM}_n(i) = f(|| \mathbf z_n - \mathbf z(i) ||) + w_n(i),
\end{equation}
where $\mathbf z_n=[x_n, y_n]^T$ is the coordinates vector of the AP and $\mathbf z(i) = [x(i), y(i)]^T$ is the coordinates vector of the device at time step~$i$. In addition, $||\mathbf z||=\sqrt{\mathbf z^T\mathbf z}$ is the $l$2-norm of a vector~$\mathbf z$. The function $f(\cdot)$ represents the distortion between true distances and measured distances.
The measurement noise $\omega_n(i)$ is assumed to be a random variable with zero mean and a standard deviation of $s_n(i)$.

Using equation (\ref{2_1}), we can estimate the distance between the $n$-th AP and the device at time step~$i$ as follows:
\begin{equation} \label{2_2}
\hat{d}_n(i) = \max \left(f^{-1}(d^{FTM}_n(i)), 0\right),
\end{equation}
where $f^{-1}(\cdot)$ is the inverse function of $f(\cdot)$ and $\max(\cdot)$ produces the largest value among all inputs.
In this paper, we call $f^{-1}(\cdot)$ the calibration curve or equation.
If the distance measurements using the FTM protocol are assumed to have only a constant offset from the true distances, the calibration equation is simply expressed by
\begin{equation} \label{2_2}
f^{-1}(d_n^{FTM}(i)) = d_n^{FTM}(i) + b,
\end{equation}
where $b$ is the distance measurement offset of the device.

However, if the distance measurements are non-linear to the true distances, a polynomial can be used to compensate for this as follows:
\begin{equation} \label{2_3}
f^{-1}(d_n^{FTM}(i)) = \sum_{l=0}^L c_l (d_n^{FTM}(i))^l,
\end{equation}
where $L$ is the highest order of the polynomial and $c_l$ is the coefficient for the $l$-th order term.
The calibration results using equation (\ref{2_2}) and (\ref{2_3}) are illustrated in Fig.~\ref{fig_ftm_scatter}(b) and (c), respectively. 
For simplicity, a second-order polynomial is used in this work.
Note that the parameters in these equations were optimally selected for generating Fig.~1, but the proposed method will estimate the parameters using unlabeled training data.

To improve the ranging accuracy, the FTM protocol also supports a burst mode that performs multiple distance measurement processes for a single ranging request and reports the average of the measured distances as a single distance estimate.
In this process, the FTM protocol can empirically obtain the standard deviation of multiple measured distances and also report this value.
Therefore, this value can be used as an estimator of standard deviation of the noise as $\hat{s}_n(i) = s_n^{FTM}(i)$, where $s_n^{FTM}(i)$ is the reported standard deviation of distance measurements between the $n$-th AP and the device at time step~$i$.

\section{AP Coordinates Estimation Method}

\begin{figure}[!t]
\centering
\includegraphics[width=0.45\textwidth]{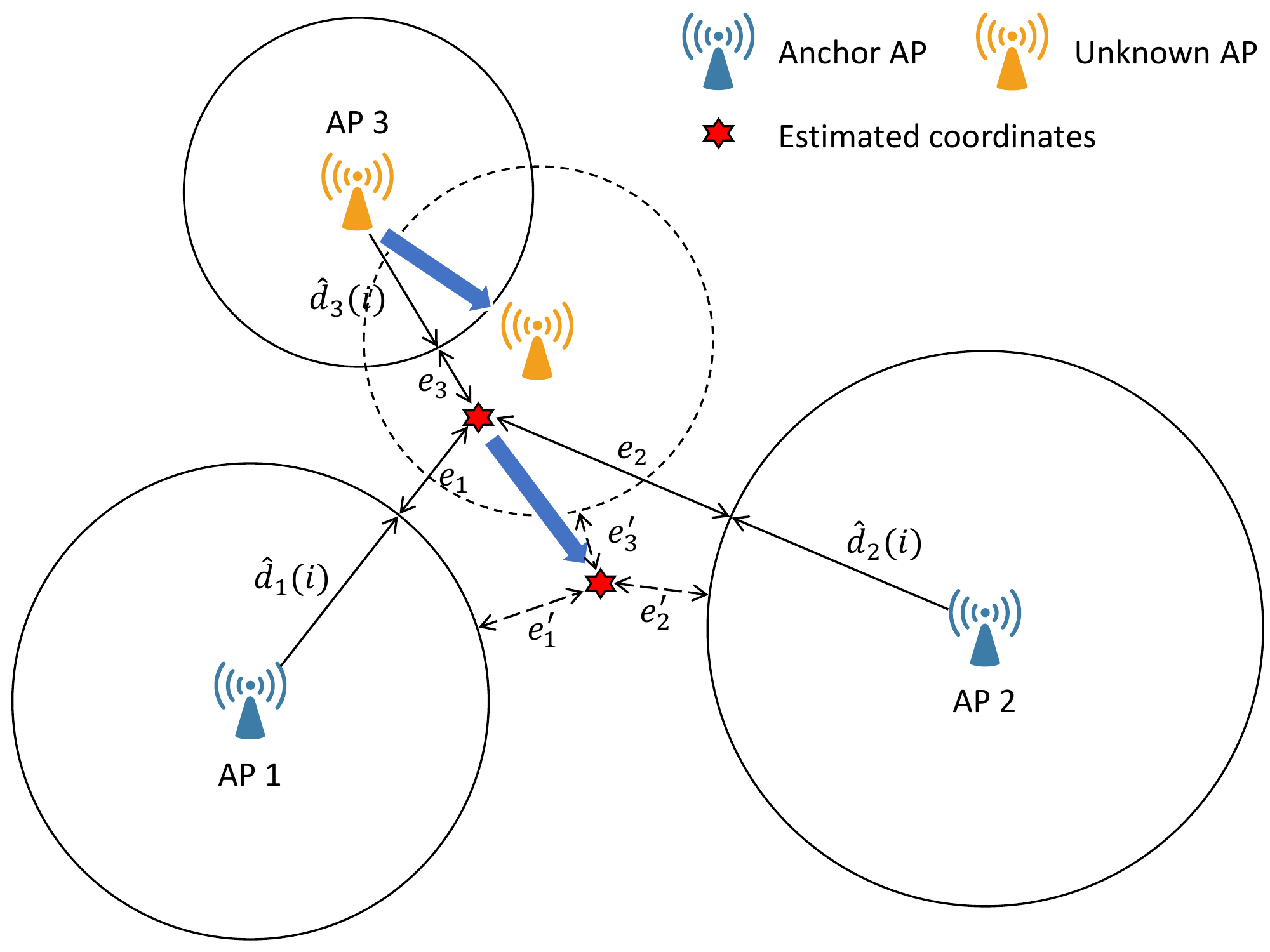}
\caption{Overview of the proposed method. If the coordinates of an unknown AP changes slightly, the estimated coordinates of the device will change and the cost functions change accordingly. The coordinates of the unknown AP are updated in the direction of reducing the cost functions.}
\label{fig_overview}
\end{figure}

\subsection{Overview}

The basic idea of the proposed method is illustrated in Fig.~\ref{fig_overview}.
Regardless of whether the coordinates of the unknown APs are correct or not (the initial coordinates of unknown APs are generally given arbitrarily), we can include all anchor and unknown APs in the location estimation phase and obtain the estimated coordinates of the device using trilateration algorithms. Subsequently, cost functions that indirectly evaluate whether current estimates of parameters (i.e., the coordinates of unknown APs and the parameters in the calibration equations) are geometrically valid can be defined.
Using these cost functions, we can analyze the impact of a slight variation of each parameter on the cost functions, and iteratively update parameters in the direction of reducing the cost functions.
For instance, the figure shows that if the estimated coordinates of AP~3 changes slightly, the estimated coordinates of the device and the cost functions will also change. Therefore, we can adjust the coordinates of AP 3 appropriately.

\subsection{Cost Functions}

Let $\hat{\mathbf z}(i)$ denote the estimated coordinates of the device using a trilateration algorithm at time step~$i$.
The estimated coordinates are generally obtained from the following information: the coordinates of the APs, distance measurements from adjacent APs, and parameters in the calibration equation.
Therefore, $\hat{\mathbf z}(i)$ is expressed as a parameterized function as follows:
\begin{equation} \label{3_2_1}
\hat{\mathbf z}(i) = \mathbf g(M(i); Z, \phi),
\end{equation}
where $M(i) = (m_1(i), ..., m_N(i))$ represents a tuple of FTM-related measurements between every AP and the device at time step~$i$. Specifically, the $n$-th element of $M(i)$ is a pair of distance and standard deviation measurements from the $n$-th AP, i.e., $m_n(i) = (d_n^{FTM}(i) , s_n^{FTM}(i))$.
In addition, $Z = (\mathbf z_1, ..., \mathbf z_N)$ represents a tuple of coordinates of all APs and $\phi$ is the set of all parameters in the calibration equation.
In equation~(\ref{3_2_1}), $M(i)$ is given as the measurement data, and $Z$ and $\phi$ are the set of trainable variables that should be appropriately optimized.

Note that some trilateration algorithms, e.g., those based on the Kalman filter~\cite{intel17, choi19}, estimate the coordinates of the device based on previous results. In this case, we can slightly modify equation (\ref{3_2_1}) as $\hat{\mathbf z}(i) = \mathbf g(M(i), \hat{\mathbf z}(i-1); Z, \phi)$ to include the previous estimate.
In addition, the FTM measurements from all APs are not always available.
Therefore, we denote $\mathcal N{(i)}\subset \mathcal N$ as the set of selected APs that are involved in the location estimation phase at time step~$i$.

A necessary condition for equation (\ref{3_2_1}) is that $\mathbf g(\cdot)$ should be differentiable with respect to all trainable variables.
Therefore, we mainly focus on linear trilateration methods such as the linear-least square (LS) or the weighted linear-least (WLS) methods that calculate the coordinates of the device using matrix operations~\cite{1275684, paula_sen_2011, 8320781}.
For the same reason, the extended Kalman filter (EKF) is also compatible with the proposed method, because it consists of matrix operations. We follow the EKF procedures introduced in~\cite{choi19}.

Irrespective of which trilateration algorithm is used, we can obtain the estimated coordinates of the device and define cost functions that indirectly evaluate the accuracy of the parameters~\cite{choi19}.
The most important cost function is the geometric cost function that is defined by
\begin{equation} \label{3_2_3}
J^{geo}(Z, \phi)=\sum_{i=1}^{T}\sum_{n\in\mathcal N{(i)}} \left(\frac{||\hat{\mathbf z}(i) - \mathbf z_n|| - \hat d_n(i)}{\hat{s}_n(i)}\right)^2.
\end{equation}
where $T$ is the total length of the measurement time steps.
This cost function measures how closely multiple circles intersect at a single point, and approaches to 0 if both the coordinates of unknown APs and distance measurements are perfect.

In addition to the geometric cost function, we can also consider other types of costs functions based on the assumption that the device cannot move far in a short time.
Therefore, the position and velocity of the device do not significantly change between two consecutive time steps. 
The cost functions related to these restrictions are given by
\begin{gather}
J^{pos}(Z, \phi) = \sum_{i=2}^T ||\hat{\mathbf z}(i) - \hat{\mathbf z}(i-1)||^2,\nonumber\\
J^{velo}(Z, \phi) = \sum_{i=3}^T ||\hat{\mathbf v}(i) - \hat{\mathbf v}(i-1)||^2,
\end{gather}
where $\hat{\mathbf v}(i) = (\hat{\mathbf z}(i) - \hat{\mathbf z}(i-1))/\Delta T$ is the estimated velocity between time step $i-1$ and $i$, and $\Delta T$ is the measurement interval.

By combining all the cost functions, we can obtain a unified cost function that is used to optimize the parameters.
This is represented by
\begin{equation} \label{3_2_7}
J(Z, \phi) = \lambda_1 J^{geo}(Z, \phi) + \lambda_2 J^{pos}(Z, \phi) + \lambda_3 J^{velo}(Z, \phi),
\end{equation}
where $\lambda_1, \lambda_2$, and $\lambda_3$ are non-negative real numbers that control the balance between cost functions.

\subsection{Optimization using the Gradient Descent Method}

Given that we define the unified cost as the function of unknown parameters in the system, we can iteratively update the parameters using the gradient descent method.
In the case that multiple mobile devices participate in collecting training data, we can further combine the cost function of each device.
Let $\mathcal K$ denote the set of $K$ mobile devices involved in the training phase.
The combined cost function is expressed by
\begin{equation}
\tilde J(Z, \Phi) = \sum_{k=1}^K J_k(Z, \phi_k), 
\end{equation}
where $J_k(Z, \phi_k)$ is the cost for the $k$-th device, obtained using equation (\ref{3_2_7}), and $\phi_k$ is the set of parameters for this user.
We denote $\Phi = \cup_{k=1}^{K} \phi_k$ as the set of all parameters for calibration.

Using the combined cost function, each trainable variable is updated using the gradient descent method as follows:
\begin{gather}
\hat{\mathbf z}_n \leftarrow \hat{\mathbf z}_n - \alpha \frac{\partial \tilde J(Z, \Phi)}{\partial \mathbf z_n}~\mbox{for}~\forall n\in \mathcal N^*,\nonumber\\
\hat{\theta} \leftarrow \hat{\theta} - \alpha \frac{\partial \tilde J(Z, \Phi)}{\partial \theta},~\mbox{for}~\forall \theta\in \Phi,
\end{gather}
where $\alpha$ is the learning rate and $\mathcal N^*\subset \mathcal N$ is the set of unknown APs. 
These partial derivatives are simultaneously evaluated with the current estimates of parameters.

\section{Experiment Results}

\begin{figure}[!t]
\centering
\includegraphics[width=0.45\textwidth]{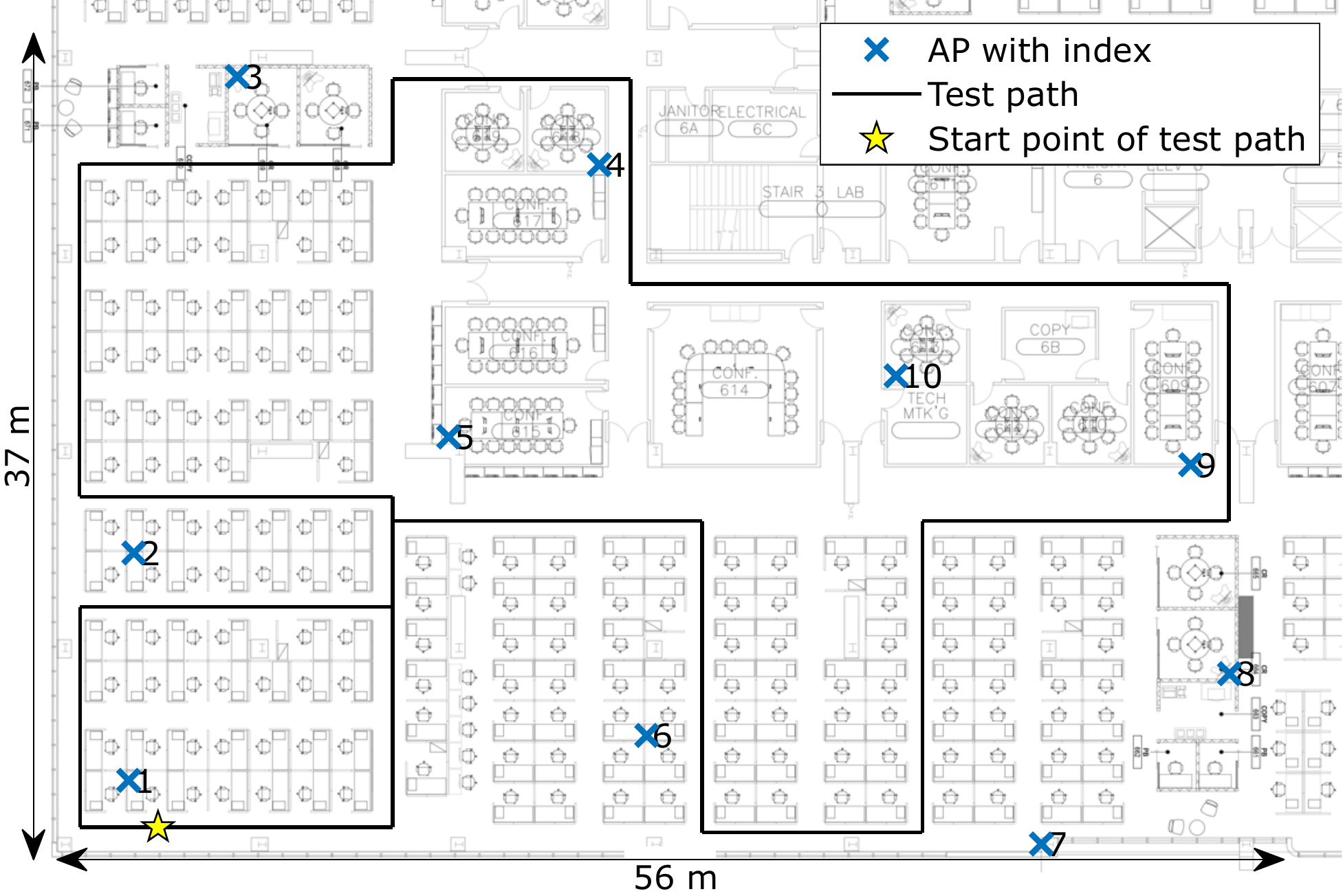}
\caption{Experiment site with 10 IEEE 802.11-2016 capable APs installed.}
\label{fig_exp_site}
\end{figure}

The performance of the proposed method was evaluated with off-the-shelf APs and devices.
Fig.~\ref{fig_exp_site} shows the floor plan of the experiment site in which 10 APs that support the FTM protocol are installed in $56\times37~\mbox{m}^2$ office environment.
These APs are equipped with the Intel Atom CPU and the Intel AC8260 Wi-Fi chipset that supports the FTM protocol.
The center frequency of each AP is  5200 or 5240~MHz (i.e., Wi-Fi channel 40 or 48) and the bandwidth is 40~MHz.
The index of each AP is also presented in the figure.
We simply chose 4 APs around the corners of the testbed area as anchor APs (i.e., AP 1, 3, 7, and 9) and the remaining 6 as unknown.

The height of each AP is 1.5~m, which is similar to the height of the mobile device.
For the mobile device, we used Google Pixel series (Pixel 1, 2, 3) that officially support the FTM protocol on Android version 9.
For each device, the training data were collected for 5 minutes by randomly walking around the testbed area and test data were obtained by following the test path shown in the figure.
The length of the test path is 235~m and it took approximately 4 minutes given that the speed of user was approximately 3~Km/h. 
While moving, each device measured the distance from all the adjacent APs every 500~ms using the FTM protocol.

For benchmark purposes, we also verify the the performance using perfect estimates of the parameters.
To this end, we utilize a second-order polynomial to calibrate raw distance measurements and optimized coefficients to minimize the mean squared error (MSE), which is defined as $MSE = E[(\hat{d} - \bar d)^2]$, where $\hat d$ and $\bar d$ represent the calibrated distance using equation (\ref{2_3}) and the true distance, respectively.
Note that test data were used to optimize parameters for benchmark scenarios because the training data do not have measured ground truth coordinates.
\emph{For this reason, the benchmark scenarios produce the best results with respect to the test data and it will be the upper bound of the proposed method}.
The optimal coefficients for each device are summarized in Table~1.

\begin{figure}[!t]
\centering
\includegraphics[width=0.45\textwidth]{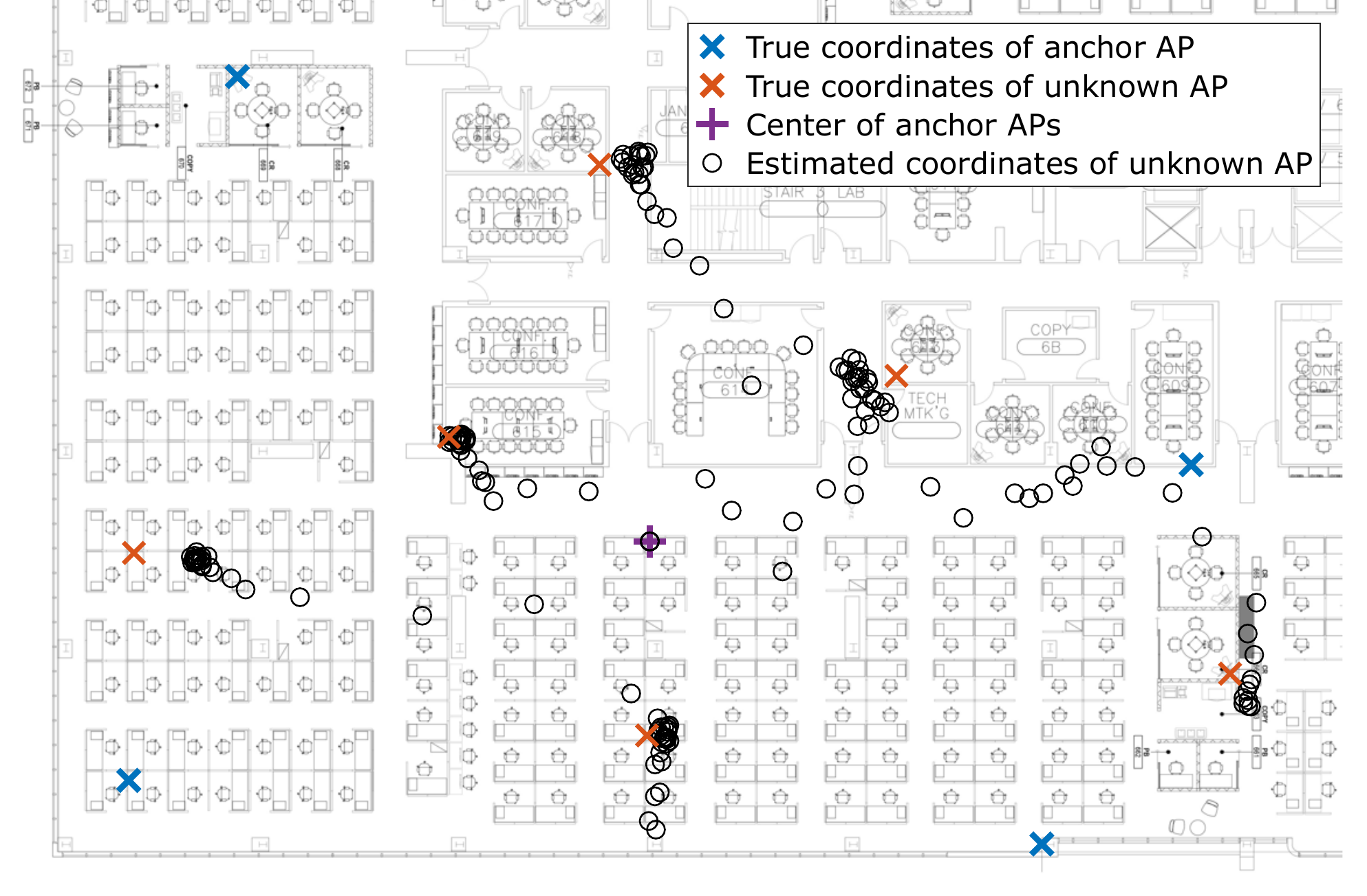}
\caption{Convergence of the coordinates of unknown APs. The estimated coordinates are presented every 20 iterations (equivalent to 1 epoch).}
\label{fig_ap_loc}
\end{figure}

The proposed method estimates the coefficients of the calibration equation while estimating the coordinates of the unknown APs.
For positioning, we exploit EKF techniques using distance measurements for up to the 5 closest APs. 
In addition, we assume $(\lambda_1, \lambda_2, \lambda_3) = (1, 0.1, 0.1)$.
Fig.~\ref{fig_ap_loc} visualizes the convergence behavior of the proposed method.
At first, the coordinates of every unknown AP are initialized as the center of the anchor APs.
As training iteration increases, the estimated coordinates approach their true coordinates. The estimated coordinates are plotted every 20 iterations.
To accelerate the convergence speed, we randomly sample only 30 consecutive time steps from training data for each device for each training iteration. Therefore, 20 iterations are equivalent to 1 training epoch.

\begin{table}[!t]
\renewcommand{\arraystretch}{1.25}
\caption{Summary of Coefficients for the Calibration Equation}
\label{table1}
\centering
\begin{tabular}{ccc}
Device &  Estimation method  & Coefficients ($c_2, c_1, c_0)$\\ \hline
\multirow{2}{*}{Pixel 1} &  Benchmark   & $(-0.0031, 0.9252, -3.8592)$\\ 
 	    &  Proposed & $(-0.0009, 0.8231, -2.3898)$\\ \hline
\multirow{2}{*}{Pixel 2}  &  Benchmark   & $(-0.0032, 0.9519, -4.0150)$\\ 
 	    &  Proposed & $(-0.0010, 0.8078, -2.0419)$\\ \hline
\multirow{2}{*}{Pixel 3}  &  Benchmark   & $(-0.0036, 0.9523, -2.7656)$\\ 
 	    &  Proposed & $(-0.0005, 0.8252, -1.3796)$\\ \hline
\end{tabular}
\end{table}

\begin{figure}
\centering
\subfloat[]{\includegraphics[width=0.24\textwidth]{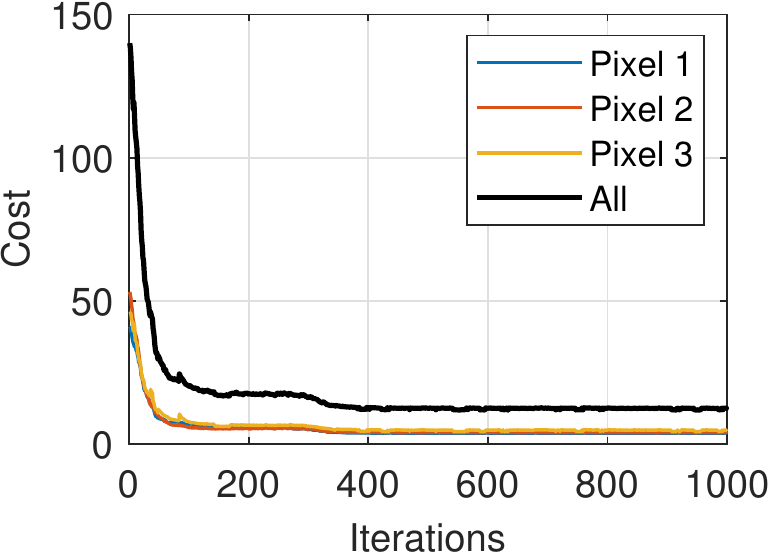}}\hfil
\subfloat[]{\includegraphics[width=0.24\textwidth]{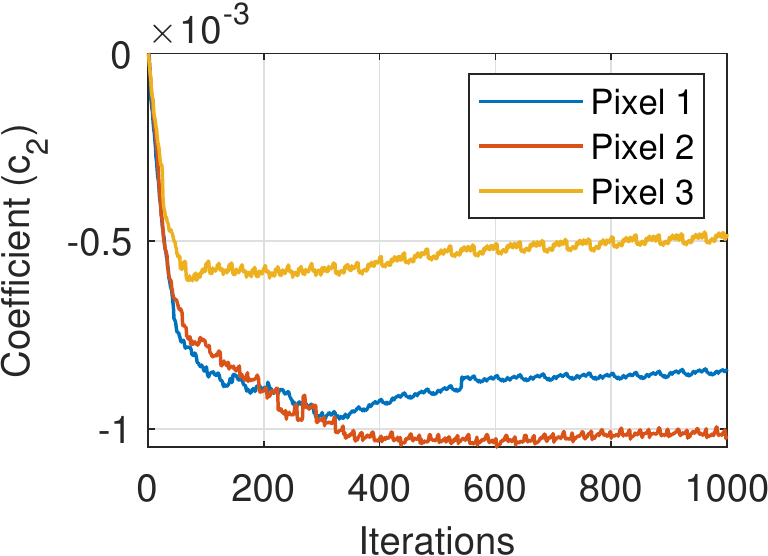}}\hfil
\subfloat[]{\includegraphics[width=0.24\textwidth]{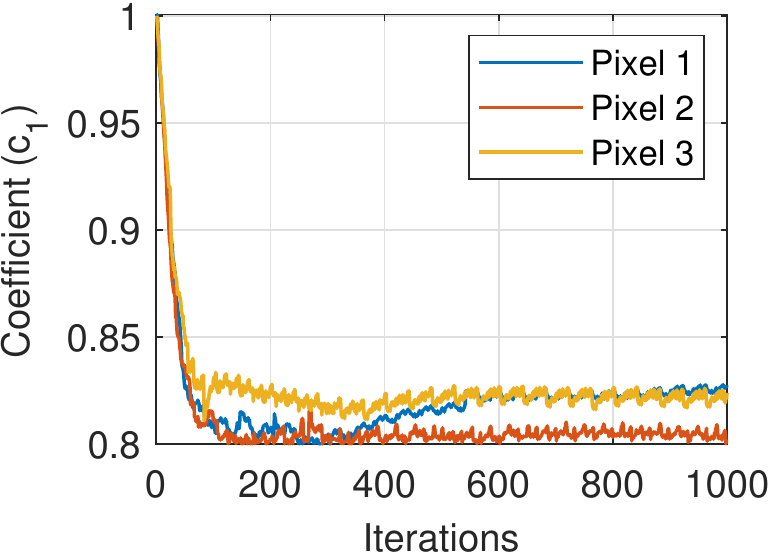}}\hfil
\subfloat[]{\includegraphics[width=0.24\textwidth]{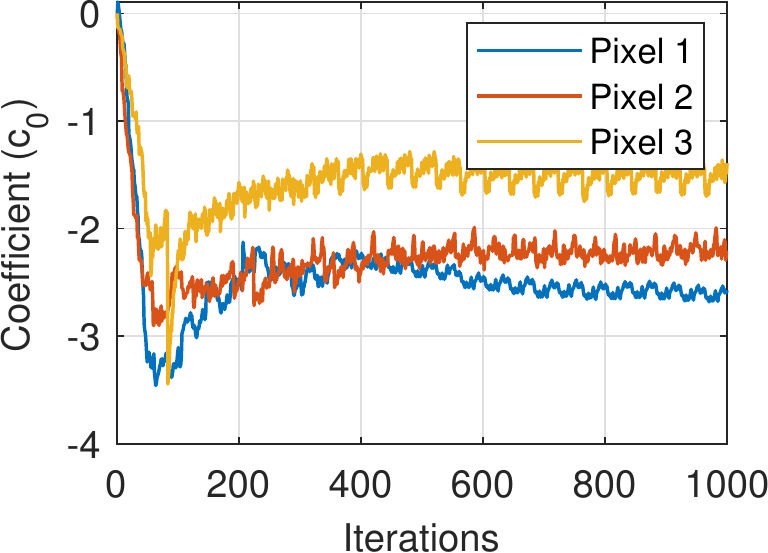}}\hfil
\subfloat[]{\includegraphics[width=0.24\textwidth]{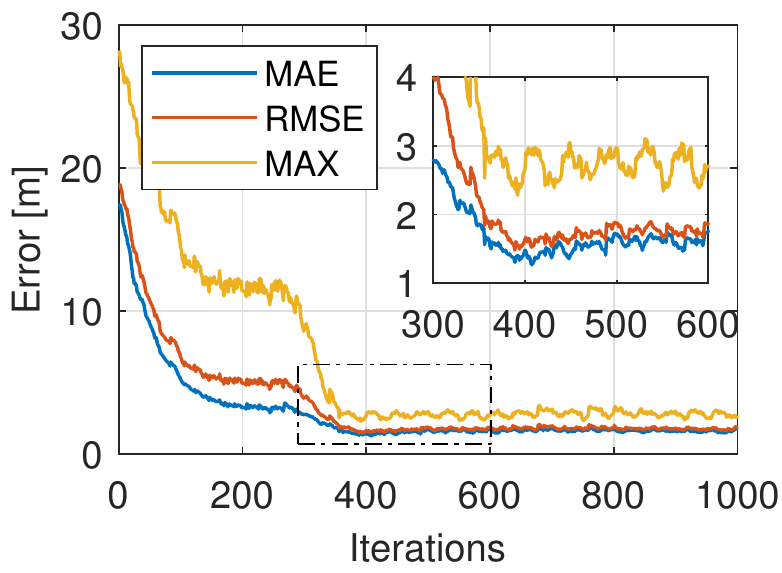}}\hfil
\subfloat[]{\includegraphics[width=0.24\textwidth]{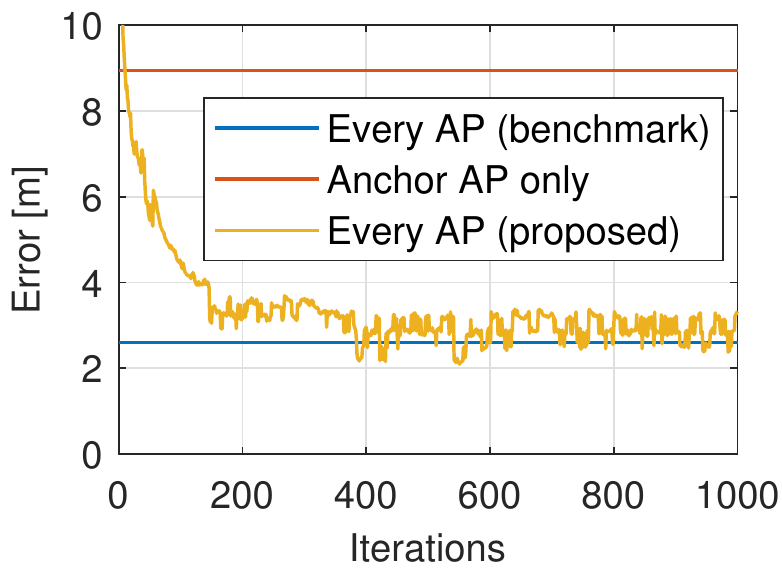}}\hfil
\caption{Training details: (a) unified cost function, (b) coefficient $c_2$, (c) coefficient $c_1$, (d) coefficient $c_0$, (e) accuracy of estimated coordinates of unknown APs, and (f) positioning accuracy with estimated AP coordinates.}
\label{fig_training}
\end{figure}

Fig.~\ref{fig_training} shows further details.
Fig.~\ref{fig_training}(a) shows that the cost of each device obtained using equation (\ref{3_2_7}) decreases with the number of iterations.
We ran 1000 training iterations for this experiment and selected parameters when the cost was minimized. 
Fig.~\ref{fig_training}(b), (c), and (d) represent the polynomial coefficients for each device. The initial coefficients were chosen as $(c_2, c_1, c_0) = (0, 1, 0)$ for each device.
The accuracy of the estimated coordinates of unknown APs is shown in Fig.~\ref{fig_training}(e). 
The performance metric used in this figure is the mean absolute error (MAE) and the root mean squared error (RMSE) that are defined as $MAE = E[||\hat{\mathbf z} - \bar{\mathbf z}||]$ and $RMSE = \sqrt{E[||\hat{\mathbf z} - \bar{\mathbf z}||^2]}$), respectively.
In this definition, $\hat{\mathbf z}$ and $\bar{\mathbf z}$ represent the estimated and the true coordinates respectively.
The maximum estimation error is also presented.
Finally, the positioning accuracy of the device using all the APs is presented in Fig.~\ref{fig_training}(f).
For performance comparison purposes, we also evaluated the positioning performance using all APs with true coordinates.
In addition, the positioning performance using only 4 anchor APs are also presented. We assume benchmark calibration for these scenarios.
The proposed method uses all APs with estimated parameters (i.e., coordinates and calibration curve).
The positioning accuracy of the proposed method closely approaches the best performance.

\begin{figure}[!t]
\centering
\includegraphics[width=0.35\textwidth]{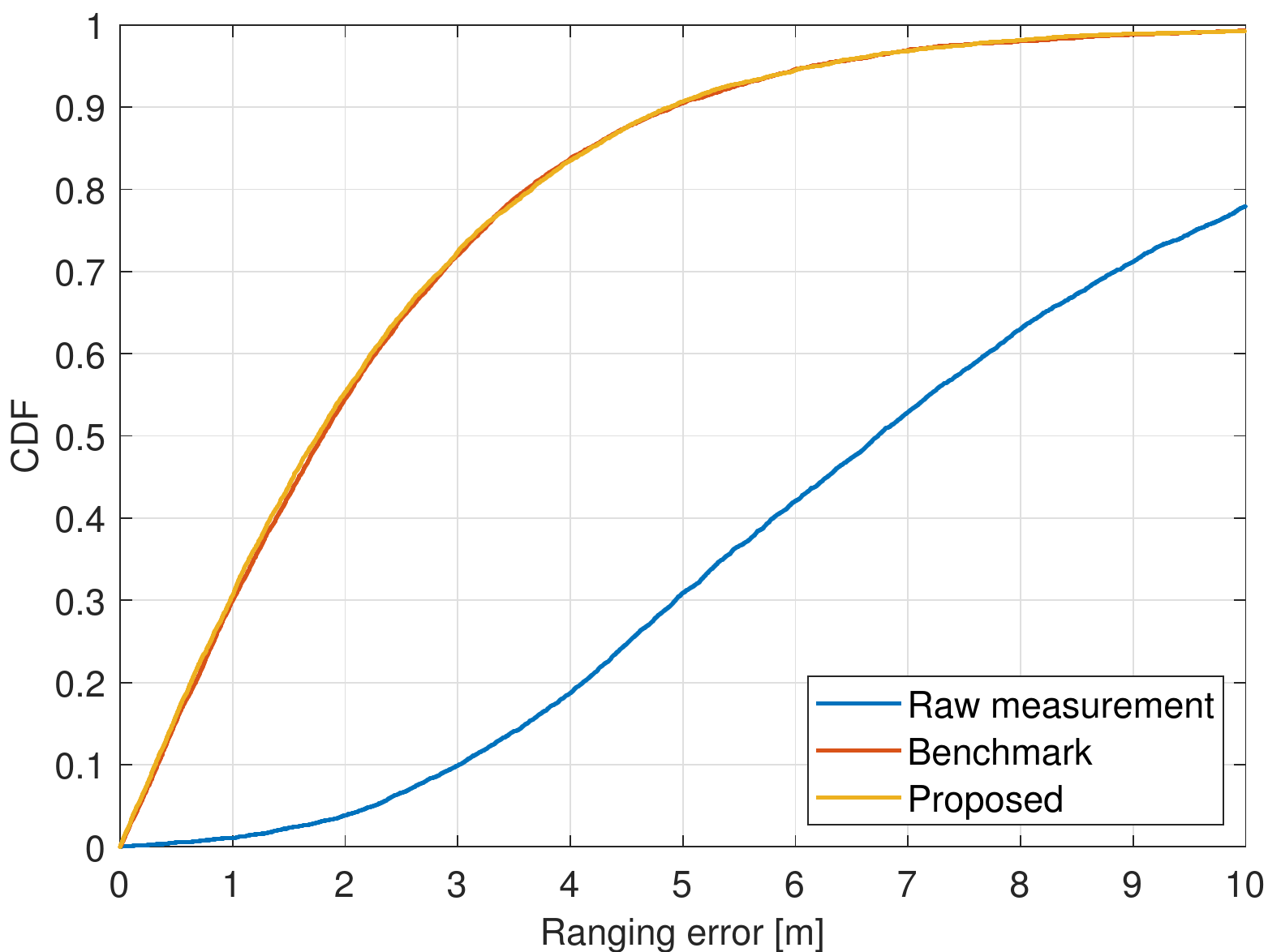}
\caption{CDF of distance estimation error for all devices. }
\label{fig_dist_cdf}
\end{figure}

\begin{figure}[!t]
\centering
\includegraphics[width=0.35\textwidth]{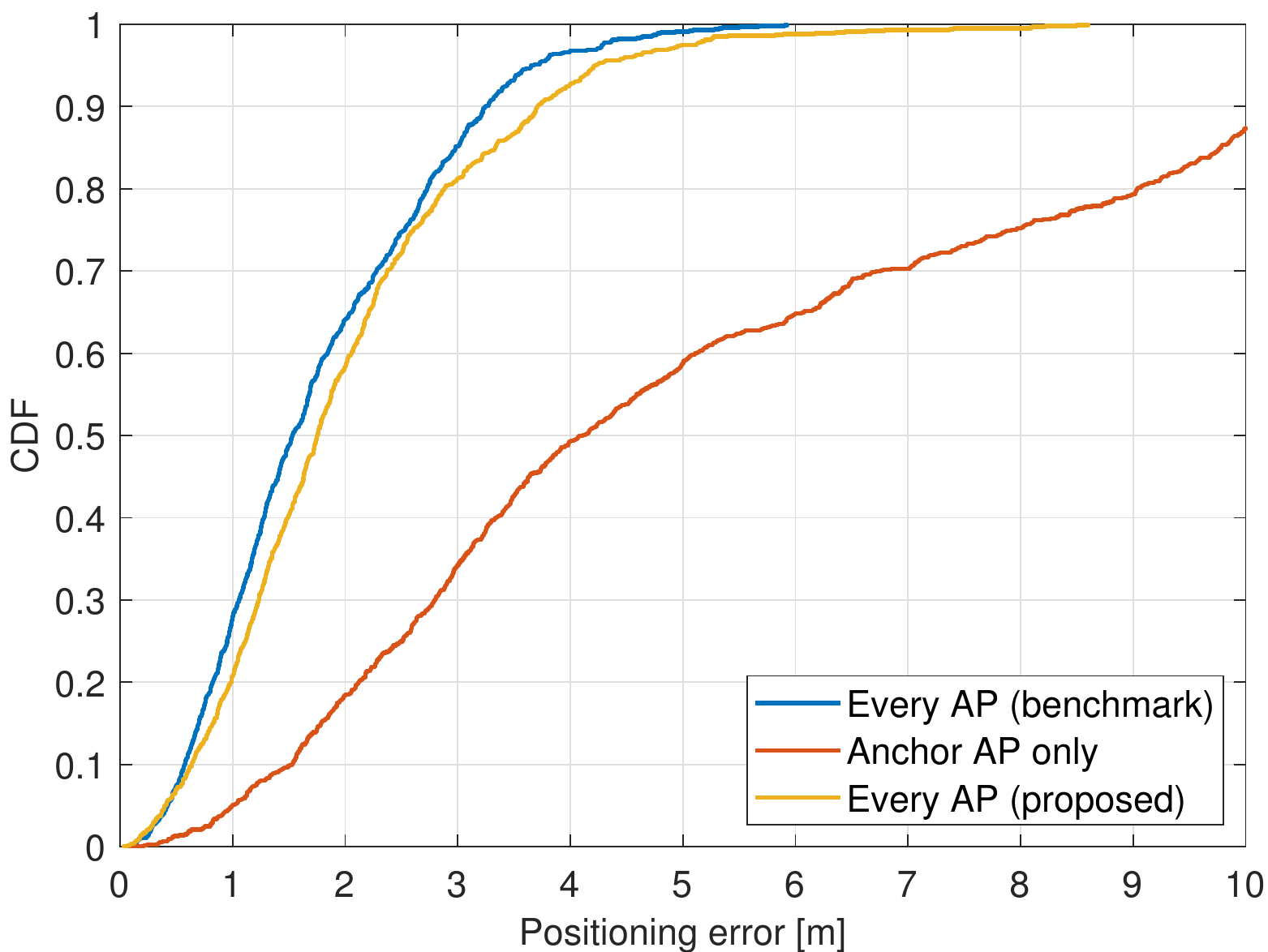}
\caption{CDF of location estimation error for all devices.}
\label{fig_loc_cdf}
\end{figure}

Fig.~\ref{fig_dist_cdf} illustrates the cumulative density function (CDF) of the distance estimation accuracy. As already shown in Fig.~\ref{fig_ftm_scatter}(a), the raw distance measurement using the FTM protocol produces significant errors.
The proposed method automatically optimizes the coefficients in the calibration equations and achieves similar results to the benchmark scenario.
Fig.~\ref{fig_loc_cdf} shows the CDF of the positioning accuracy.
Even though well-calibrated distance measurements are used, the 4 anchor nodes are not sufficient to provide meaningful results.

\begin{figure}[!t]
\centering
\includegraphics[width=0.45\textwidth]{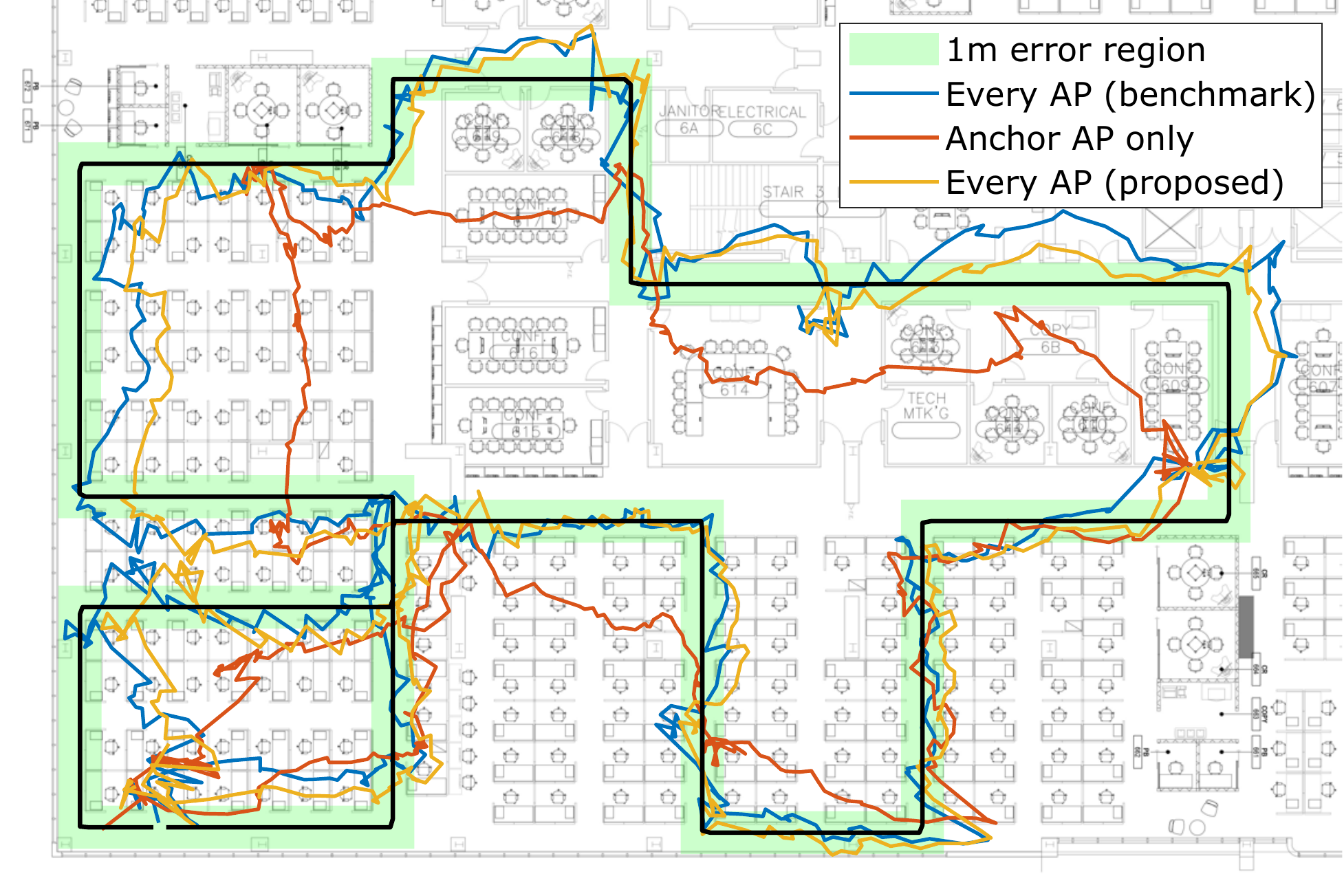}
\caption{Estimated trajectory with the Google Pixel 3.}
\label{fig_trj}
\end{figure}

Finally, Fig.~\ref{fig_trj} represents the estimated trajectory of the Pixel~3 device for different scenarios.
The area in green represents a 1~m error region, which means that every point in the region is at most 1~m from the true path.
Similar to the previous results, using only 4 anchor nodes cannot produce an accurate trajectory of the device. 
By securing for more anchor nodes in an unsupervised manner, the proposed method is able to produce accurate positioning results.

\section{Conclusion}

In this paper, we proposed an unsupervised learning technique to estimate unknown parameters in a system, including the coordinates of unknown APs.
By simply moving around an area where unknown APs coexist with anchor APs, the proposed method automatically determined the coordinates of unknown APs as well as an appropriate calibration curve for each device.
Using the proposed algorithm, a greater number of anchor nodes can be secured for positioning purposes.
Therefore, high accuracy positioning results can be obtained using dense anchor nodes.
The proposed method can be applied in numerous ways.
For instance, a service operator can temporarily deploy a few APs at the reference locations for which the coordinates are easily obtainable (e.g., corners of the building) to estimate the coordinates of unknown APs.

\begin{thebibliography}{10}
\providecommand{\url}[1]{#1}
\csname url@samestyle\endcsname
\providecommand{\newblock}{\relax}
\providecommand{\bibinfo}[2]{#2}
\providecommand{\BIBentrySTDinterwordspacing}{\spaceskip=0pt\relax}
\providecommand{\BIBentryALTinterwordstretchfactor}{4}
\providecommand{\BIBentryALTinterwordspacing}{\spaceskip=\fontdimen2\font plus
\BIBentryALTinterwordstretchfactor\fontdimen3\font minus
  \fontdimen4\font\relax}
\providecommand{\BIBforeignlanguage}[2]{{%
\expandafter\ifx\csname l@#1\endcsname\relax
\typeout{** WARNING: IEEEtran.bst: No hyphenation pattern has been}%
\typeout{** loaded for the language `#1'. Using the pattern for}%
\typeout{** the default language instead.}%
\else
\language=\csname l@#1\endcsname
\fi
#2}}
\providecommand{\BIBdecl}{\relax}
\BIBdecl

\bibitem{Wang2003AnIW}
Y.-C. Wang, X.~Jia, and H.~K. Lee, ``An indoor wireless positioning system
  based on wireless local area network infrastructure,'' in \emph{Proc. 6th
  International Symposium on Satellite Navigation Technology Including Mobile
  Positioning and Location Services}, 2003.

\bibitem{5425237}
J.~{Yang} and Y.~{Chen}, ``Indoor localization using improved rss-based
  lateration methods,'' in \emph{Proc. IEEE Global
  Telecommunications Conference (GLOBECOM)}, Nov. 2009, pp. 1--6.

\bibitem{5766644}
B.~{Kim}, W.~{Bong}, and Y.~C. {Kim}, ``Indoor localization for Wi-Fi devices
  by cross-monitoring AP and weighted triangulation,'' in \emph{Proc. IEEE
  Consumer Communications and Networking Conference}, Jan. 2011.
  
\bibitem{6488558}
Y.~{Wang}, X.~{Yang}, Y.~{Zhao}, Y.~{Liu}, and L.~{Cuthbert}, ``Bluetooth positioning using RSSI and
  triangulation methods,'' in \emph{Proc. IEEE Consumer Communications and
  Networking Conference}, Jan. 2013. 
  
  
\bibitem{3gppSCM}
``Spatial channel model for multiple input multiple output ({MIMO})
  simulations,'' 3GPP TR25.996 release 11, Sep. 2012.

\bibitem{3gpp3DSCM}
``Study on {3D} channel model for {LTE},'' 3GPP TR36.873 release 12, Sep. 2014.

\bibitem{832252}
P.~{Bahl} and V.~N. {Padmanabhan}, ``RADAR: An in-building RF-based user
  location and tracking system,'' in \emph{Proc. IEEE Conference on Computer Communications (INFOCOM)}, Mar. 2000.
  
\bibitem{1047316}
P.~{Prasithsangaree}, P.~{Krishnamurthy}, and P.~{Chrysanthis}, ``On indoor
  position location with wireless LANs,'' in \emph{Proc. 13th IEEE International
  Symposium on Personal, Indoor and Mobile Radio Communications (PIMRC)}, Sep. 2002.
  
\bibitem{horus05}
M.~Youssef and A.~Agrawala, ``The Horus WLAN location determination system,''
  in \emph{Proc. 3rd International Conference on Mobile Systems,
  Applications, and Services (MobiSys)}, 2005.
  

\bibitem{6550414}
L.~{Zhang}, X.~{Liu}, J.~{Song}, C.~{Gurrin}, and Z.~{Zhu}, ``A comprehensive
  study of bluetooth fingerprinting-based algorithms for localization,'' in
  \emph{Proc. 27th International Conference on Advanced Information Networking
  and Applications Workshops}, Mar. 2013.
  
\bibitem{7438932}
X.~{Wang}, L.~{Gao}, S.~{Mao}, and S.~{Pandey}, ``CSI-based fingerprinting for
  indoor localization: A deep learning approach,'' \emph{IEEE Trans. on
  Veh. Technol.}, vol.~66, no.~1, pp. 763--776, Jan. 2017.

\bibitem{intel17}
L.~Banin, O.~Bar-Shalom, N.~Dvorecki, and Y.~Amizur, ``High-accuracy indoor
  geolocation using collaborative time of arrival (CToA),'' \emph{Intel White
  Paper}, Sep. 2017.

\bibitem{Ibrahim:2018:VAE:3241539.3241555}
M.~Ibrahim, H.~Liu, M.~Jawahar, V.~Nguyen, M.~Gruteser, R.~Howard, B.~Yu, and
  F.~Bai, ``Verification: Accuracy evaluation of WiFi fine time measurements on
  an open platform,'' in \emph{Proc. 24th Annual International
  Conference on Mobile Computing and Networking (MobiCom)}, 2018.
  
\bibitem{choi19}
J.~Choi, Y.-S. Choi, and S.~Talwar, ``Unsupervised learning techniques for
  trilateration: From theory to android App implementation,'' \emph{Submitted
  to IEEE Access}, 2019.

\bibitem{8053692}
V.~{Djaja-Josko}, ``A new anchor nodes position determination method supporting
  UWB localization system deployment,'' in \emph{Signal Processing
  Symposium (SPSympo)}, Sep. 2017.
  
\bibitem{Costa:2006:DWS:1138127.1138129}
J.~A. Costa, N.~Patwari, and A.~O. Hero, III, ``Distributed
  weighted-multidimensional scaling for node localization in sensor networks,''
  \emph{ACM Trans. Sen. Netw.}, vol.~2, no.~1, pp. 39--64, Feb. 2006.
  
\bibitem{8439937}
W.~{Yu}, J.~{Choi}, Y.~{Kim}, W.~{Lee}, and S.~{Kim}, ``Self-organizing
  localization with adaptive weights for wireless sensor networks,'' \emph{IEEE
  Sensors J.}, vol.~18, no.~20, pp. 8484--8492, Oct. 2018.

\bibitem{ftm_cal}
\BIBentryALTinterwordspacing
 [Online]. Available:
  \url{https://source.android.com/devices/tech/connect/wifi-rtt}
\BIBentrySTDinterwordspacing

\bibitem{1275684}
K.~W. Cheung, H.~C. So, W.~. Ma, and Y.~T. Chan, ``Least squares algorithms for
  time-of-arrival-based mobile location,'' \emph{IEEE Trans. Signal
  Process.}, vol.~52, no.~4, pp. 1121--1130, April 2004.

\bibitem{paula_sen_2011}
B.~A.~M. Tarrío, P. and J.~R. Casar, ``Weighted least squares techniques for
  improved received signal strength based localization,'' \emph{Sensors}, vol.
  11(9), Sep. 2011.

\bibitem{8320781}
K.~Bregar and M.~Mohorčič, ``Improving indoor localization using
  convolutional neural networks on computationally restricted devices,''
  \emph{IEEE Access}, vol.~6, pp. 17,429--17,441, 2018.

\end{thebibliography}


\end{document}